\documentclass[10pt,conference]{IEEEtran}
\usepackage[T1]{fontenc}
\usepackage{amsmath,amssymb}
\usepackage{booktabs,array,tabularx}
\usepackage{graphicx,xcolor}
\usepackage{enumitem}
\setlist[itemize]{nosep,leftmargin=*}
\setlist[enumerate]{nosep,leftmargin=*}
\usepackage[most]{tcolorbox}
\definecolor{rqback}{HTML}{F2F4F7}
\definecolor{rqframe}{HTML}{4A5568}
\usepackage{xurl}
\usepackage{cite}
\usepackage{balance}
\usepackage[hidelinks]{hyperref}
\newcommand{\tracer}{\textsc{Tracer}}
\newcommand{\code}[1]{\texttt{\detokenize{#1}}}
\newcommand{\rqfinding}[3]{%
  \begin{tcolorbox}[
    colback=rqback, colframe=rqframe,
    boxrule=0.4pt, leftrule=3.5pt, arc=1.5pt,
    left=6pt, right=6pt, top=3pt, bottom=3pt,
    fontupper=\small
  ]
  \textbf{Answer to #1.} 
  #3
  \end{tcolorbox}%
}
\title{Securing Agent-Generated Code \\with Its Own Development History}

\author{
\IEEEauthorblockN{Prateek Kumar Rajput\IEEEauthorrefmark{1},
Shahzeb Khan\IEEEauthorrefmark{2},
El-Hacen Diallo\IEEEauthorrefmark{1},\\
Yewei Song\IEEEauthorrefmark{1},
Jacques Klein\IEEEauthorrefmark{1}, and
Tegawend\'e F. Bissyand\'e\IEEEauthorrefmark{1}}
\IEEEauthorblockA{\IEEEauthorrefmark{1}University of Luxembourg, Luxembourg\\
\{prateek.rajput, el-hacen.diallo, yewei.song, jacques.klein, tegawende.bissyande\}@uni.lu}
\IEEEauthorblockA{\IEEEauthorrefmark{2}Nostrum Biodiscovery S.L., Barcelona, Spain\\
shahzeb.khan@nostrumbiodiscovery.com
}
}

\begin{document}
\bstctlcite{IEEEexample:BSTcontrol}
\maketitle
\begin{abstract}
Pull requests written by coding agents can pass a project's test suite while omitting a guard or a validation step. To find the missing condition, a reviewer currently relies on static analysis of the final code. An agent-generated change, however, is accompanied by a development history that a human commit does not provide. The agent's trajectory records each edit together with the agent's stated assumptions, and structural differencing shows how each changed function evolved. We present \tracer{}, a workflow that uses this development history as evidence for security review. \tracer{} tracks changes at the symbol level, treats the agent's recorded rationale as claims to be verified against the current code, accepts a repair only after assessment and functional validation, and attaches to each accepted repair an executable check that later changes must satisfy. On the SusVibes benchmark, symbol-level tracking places the vulnerable symbol in the review scope for 94.6\% of agent patches, compared with 13.4\% for the primary scanner warning. With two investigator models, the configuration that includes the agent's rationale yields the largest number of verified security repairs while preserving functionality, and adding AST history to it does not increase that number. A funnel analysis identifies where the remaining candidates are lost. In a study of structural-biology software, the combination of repair memory and a boundary check allows a subsequent feature request to succeed in nine of ten tasks without reintroducing the repaired vulnerability, whereas functional feedback alone succeeds in four and reintroduces the vulnerability in three.
\end{abstract}
\begin{figure*}[t]
\centering
\includegraphics[width=0.70\textwidth]{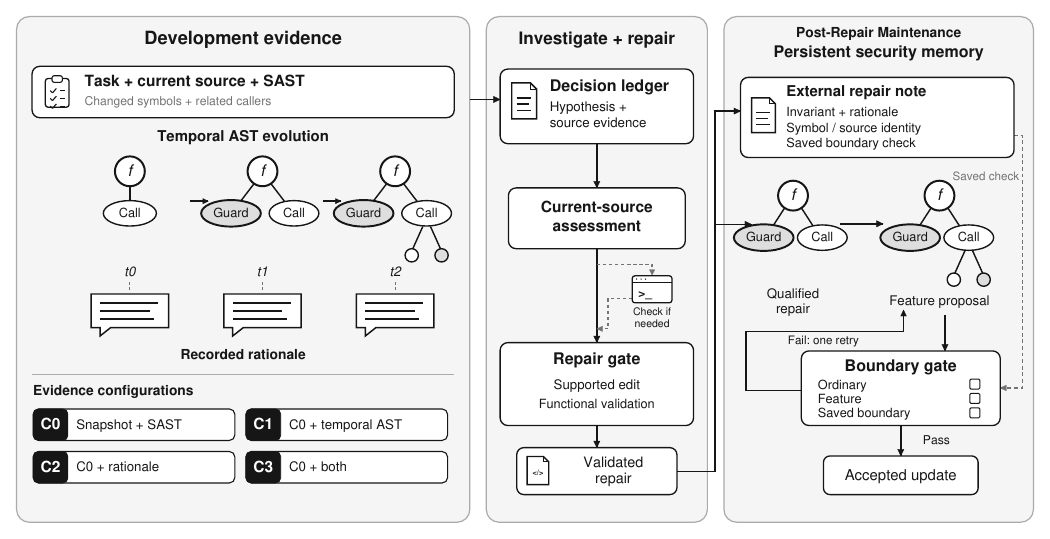}
\caption{\tracer{} links symbol-level development evidence to guarded security repair, then preserves repair invariants and executable checks for subsequent feature evolution. C0 to C3 vary the investigation evidence.}
\label{fig:architecture}
\end{figure*}

\section{Introduction}
Coding agents now navigate repositories, edit source files, and run test suites through interfaces such as SWE-agent~\cite{sweagent}, and agent-based repair systems apply the same tool-driven loop to debugging~\cite{repairagent}. A patch that passes the project's tests may still be insecure. Studies of generated code and of developers working with coding assistants report recurring security weaknesses~\cite{asleep,perry,copilotweaknesses}, and SusVibes measures the gap at repository scale by grading functionality and security with separate test suites~\cite{susvibes,susvibesartifact}. In the cohort studied in this paper, 350 of 354 agent patches pass their functional tests and 69 pass their security tests. A team that merges a functionally correct patch therefore merges the missing security condition with it.

Reviewing such a patch requires reconstructing what the agent assumed and which condition it omitted. We refer to this reconstruction as \emph{code archaeology}. Static application security testing (SAST) tools such as CodeQL~\cite{codeql} and Semgrep~\cite{semgrep} report locations and flows in the final snapshot, and industrial deployments have established that diagnostics must be actionable at review time to be used~\cite{johnson,googleanalysis,infer}. LLM-assisted taint and dataflow analysis~\cite{iris,llmdfa,repoaudit}, analyzer-guided repair~\cite{inferfix}, and analysis-guided proof-of-concept generation~\cite{pagent} connect model judgments to source and execution evidence. All of these techniques operate on the current source. They do not explain why the agent omitted a guard or what it assumed about upstream inputs, and in our cohort the primary SAST warning is located on the symbol touched by the reference security fix in 13.4\% of patches. In the remaining cases, the reviewer must first determine which symbol to examine.

An agent-generated change is accompanied by evidence that a human commit does not provide. The trajectory records each edit, each tool call, and the agent's ReAct-style explanation of its actions~\cite{react}. Structural differencing~\cite{gumtree} shows how each changed function evolved, for instance a guard added at one turn and removed at a later one. The recorded rationale states assumptions about inputs, callers, and libraries. These statements are not faithful accounts of the model's internal reasoning~\cite{turpin,lanham}, and we treat them as claims to be verified. A claim that an input is already validated, for example, creates an obligation to inspect the validator and its callers. The thesis of this paper is that this development history constitutes security evidence, and that review of agent-generated code should be organized around the changed symbols and the recorded claims in addition to scanner warnings.

A second problem is that repairs are not preserved over time. Security fixes are often committed without being labelled as such~\cite{colefunda}, and a later change can undo them~\cite{undoingfix}. Section~\ref{sec:motivation} describes an observed case in which a request for legacy checkpoint compatibility re-enabled the unrestricted deserialization that an earlier repair had removed, while every functional test passed. The evidence gathered during investigation, that is the invariant, its rationale, and an executable boundary check, can remain attached to the repaired symbol and be evaluated against every later proposal.

We present \tracer{} (Trace-based Runtime Analysis for Continuous Evolution and Release), a workflow that addresses both problems and is packaged as a portable skill that a coding agent or a continuous-integration pipeline can load. Given a Git range or an agent trajectory, \tracer{} tracks changes at symbol level, keeps typed AST evolution and recorded rationale as separate evidence channels, and admits a repair only through a gate that requires a supported assessment of the current source, review of downstream consumers, and public functional validation. We call this \emph{guarded} investigation. After a repair, \tracer{} writes a note that binds the invariant to the symbol's structural hash and to an independently owned boundary check that later changes must pass.

We evaluate \tracer{} in two settings. On SusVibes, 387 agent patches from seven coding generators are reviewed in four evidence configurations by two investigator models, DeepSeek-V3.2 and Gemini-3.8-Flash, for 3,096 repair runs. Symbol-level tracking brings the reference-vulnerable symbol into the review scope for 94.6\% of patches, against 13.4\% for the primary SAST location. Adding the agent's recorded rationale gives the highest verified security count with both investigators, 76 and 75 passes from 68 on the 345 patches evaluated in every configuration, with at most one functional regression. The gain is modest and its confidence interval includes zero, so we report it as an ordering that holds across two independent investigators rather than as a large effect. Adding AST history to the rationale configuration lowers the count with both investigators. In a second study, five components of structural-biology software relevant to the in silico drug discovery pipeline of Nostrum Biodiscovery, our industrial partner~\cite{rfdiffusion,proteinmpnn,haddock3,dockq,vina,foldseek}, are repaired and then followed through ten feature-evolution tasks under a one-retry budget. Functional feedback alone delivers both the feature and the repaired property in 4 of 10 tasks and accepts three contract violations. Repair memory with a boundary gate reaches 9 of 10 and accepts none.

The paper makes three contributions.
\begin{itemize}
\item A provenance-aware workflow and a portable skill. \tracer{} connects symbol-level change tracking, separate AST-evolution and rationale channels, guarded static and dynamic investigation, and persistent repair memory, and runs alongside an agent-generated change in an agent host or in CI.
\item A two-investigator evaluation on 387 agent patches. The SusVibes study compares four evidence configurations across seven coding generators, reporting review coverage, verified security outcomes, and functional preservation together. The structural-biology study isolates repair memory from boundary feedback in a 2$\times$2 factorial under silent-fix feature evolution.
\item A four-step repair funnel and a failure analysis. Following every attempt from hypothesis to retained patch separates losses in diagnosis, turn scheduling, and semantic repair from operational errors, and yields design lessons for agentic review systems.
\end{itemize}

\section{Background and Motivation}
\label{sec:motivation}
This section states the review problem that \tracer{} addresses, illustrates it with a case from the industrial study, and describes the two evidence channels that an agent trajectory makes available.

\subsection{Code Archaeology in Agent-Generated Pull Requests}
When an agent-generated pull request works, the reviewer still has to decide which implementation assumptions deserve scrutiny and which security conditions must survive later maintenance. CodeQL and Semgrep report locations and flows in the final snapshot~\cite{codeql,semgrep}, and connecting those findings to the submitted change, its callers, and the intended contract is left to the reviewer.

The industrial study provides a concrete instance. Nostrum Biodiscovery, a spin-off of IRB Barcelona and the Barcelona Supercomputing Center, provides in silico drug discovery and enzyme engineering services built on molecular simulation, and the structure-prediction, design, and docking tools examined in Section~\ref{sec:practice} are relevant to that pipeline. These tools load model checkpoints, parse structure files, extract archives, and write output paths, and each of these operations is a boundary that a compatibility change can weaken. The following case from that study illustrates both the review problem and the maintenance problem. RFdiffusion's \code{Sampler.load_checkpoint} loaded PyTorch checkpoints with unrestricted pickle deserialization, which under PyTorch 2.5.1 permits code execution from an untrusted checkpoint~\cite{pytorchserialization,pythonpickle}. The repair sets \code{weights_only=True} with a scoped \code{safe_globals} allowlist. A later feature request asked for compatibility with legacy \code{argparse.Namespace} configurations. With functional feedback only, DeepSeek-V3.2 implemented the compatibility branch with \code{weights_only=False} inside the \code{safe_globals} context, and all ordinary and feature tests passed. The test suite does not record the reason for the restriction. With the repair note and the saved boundary check, the same model extended the allowlist to \code{argparse.Namespace} and retained restricted loading. Section~\ref{sec:practice} reports the full factorial. Preserving review evidence as reusable checks is also what secure-development guidance asks for when it calls for retaining vulnerability knowledge across releases~\cite{ssdf}.

\subsection{Evidence Available in an Agent Trajectory}
Two channels of the trajectory carry information that the final snapshot does not. The first is structural. Source lines shift, definitions are added or renamed, and refactoring moves function boundaries, so a line-based \code{git diff} does not say which function-level contract changed, and intermediate fragments may not parse. CoLeFunDa's GumTree-based function-change descriptions~\cite{colefunda} are the model for our compact summaries of symbol edits, which avoid carrying full ASTs across versions~\cite{gumtree,gumtree3}. The second channel is the recorded reasoning. ReAct-style trajectories interleave explanations with actions~\cite{react}, so the task description, the stated assumptions, and the observed edits can be read together. Neither channel supplies the requested behavior or the domain assumptions behind the code, and the explanations are auditable claims rather than guaranteed accounts of the model's reasoning~\cite{turpin,lanham}. A function can run correctly while trusting a caller without justification or implementing only part of the task's contract.

\section{TRACER Workflow and Architecture}
\label{sec:architecture}
Figure~\ref{fig:architecture} gives an overview of the workflow. \tracer{} builds a symbol-level view of the change with scanner evidence attached, adds the two history channels, drives investigation and repair through a decision ledger and a repair gate, and emits the persistent note and boundary check that later maintenance is tested against.

\subsection{Symbol-Level Change Tracking and Static Analysis}
\tracer{} indexes code at the symbol level with the standard-library \code{ast.parse} parser~\cite{pythonast} instead of reviewing raw diff lines. For each snapshot, the index records functions, methods, and classes by repository path, qualified name, line span, and structural hash. When a patch is submitted, \tracer{} maps modified lines to their enclosing symbols, excluding test and documentation files. It runs Semgrep~\cite{semgrep} and CodeQL~\cite{codeql} on the snapshot and attaches warnings and flow evidence to the affected symbols. Guard conditions and possible callers in the same file are collected as context, and scanner evidence keeps its tool attribution and its link to the changed function.

\subsection{Temporal AST Evolution via GumTree}
To show how a modified symbol changed during development, \tracer{} parses before-and-after source pairs into typed Python ASTs and computes structural edit scripts with GumTree 3.0.0~\cite{gumtree,gumtree3}. It restricts the operations to the target symbols and highlights added, modified, or removed guards and calls. Whether the suspected boundary is enforced is always decided on the current code. If a fragment fails to parse or a symbol cannot be linked across versions, \tracer{} records the gap rather than assuming continuity through a rename or a refactor.

\subsection{Symbol-Linked Reasoning Traces}
Alongside the task description and the code changes, \tracer{} extracts the coding agent's original ReAct-style explanations~\cite{react}, excluding fenced code blocks so that natural-language reasoning stays separate from source edits. Excerpts are linked to symbols through recorded file paths, symbol mentions, and matching edit fragments. Because stated rationale does not always match what the code does~\cite{turpin,lanham}, each explanation is treated as a hypothesis to test. When an explanation asserts an assumption about inputs or callers, the quotation and a verification plan are attached to the target symbol, and the investigator checks whether the current source supports or contradicts it. The quotation stays attached to its assessment so that a later reviewer can inspect both.

\subsection{Decision Ledger, Repair Gate, and Post-Repair Maintenance}
The investigator records each security hypothesis, its target symbol, input boundary, and suspected flaw in a decision ledger saved as JSON. It checks static warnings against the current source and reads whatever historical evidence its configuration provides. When the evidence is inconclusive, \tracer{} runs targeted Python checks in the analysis-guided testing pattern of PAGENT~\cite{pagent}, executing the target on ordinary and boundary inputs to observe whether the suspected flaw can be triggered. The repair gate accepts a candidate only after a supported assessment, a review of the relevant downstream consumers, and public functional validation.

After a repair, \tracer{} saves a persistent note that binds the security invariant to the repaired symbol's hash and to an independently owned executable boundary check. Later feature proposals are checked against the ordinary tests, the new feature tests, and the saved boundary check. On failure, the maintenance gate returns the failing check as feedback for one bounded retry, and otherwise retains the secure starting code. The portable \code{tracer-security} skill\footnote{\url{https://github.com/pkrajput/TRACER_skill}} packages this workflow for host coding agents. The host supplies editing and test interfaces, and CI can run the saved checks on future commits without the benchmark harness.

\section{Experimental Setup}
\label{sec:design}
This section states the research questions, describes the SusVibes cohort and the two investigator models, defines the four evidence configurations, and fixes the evaluation protocol.

\subsection{Research Questions}
Five research questions follow an agent-generated change from localization to later maintenance.

\noindent\textbf{RQ1 (Localization Coverage).} \emph{How often does symbol-level change tracking bring the vulnerable symbol into the review scope, compared with stock SAST locations?}

\noindent\textbf{RQ2 (End-to-End Verified Repair).} \emph{Does guarded investigation improve verified security outcomes without degrading functionality?}

\noindent\textbf{RQ3 (Provenance Channels).} \emph{How do temporal AST history, recorded rationale, and their combination affect verified repair?}

\noindent\textbf{RQ4 (Repair Funnel Bottlenecks).} \emph{Where are candidates lost between symbol localization and verified repair?}

\noindent\textbf{RQ5 (Persistent Memory Under Silent-Fix Evolution).} \emph{Can repair memory and boundary gates preserve a repaired property while a subsequent feature change succeeds?}

\subsection{The SusVibes Benchmark and Study Dataset}
We use the 186-task SusVibes v1.0 release. The original paper reports statistics over 200 tasks~\cite{susvibes,susvibesartifact}, with repositories averaging 162K lines across 867 files, 108 projects, 77 CWE categories, and on average 68.8 functional tests and 3.7 security tests per task, so agents work against existing interfaces and callers rather than isolated functions.

Table~\ref{tab:inventory} lists 1,302 archived patches from seven coding generators across the 186 Python tasks, taken from the public SWE-agent trajectories and patches on the SusVibes leaderboard~\cite{susvibestrajectories}. Of these, 472 carry a publisher-reported functional success. Source screening and local environment reproduction admit 387 patches across 118 tasks. Original evaluations complete for 354 of them, of which 350 pass the functional tests and 69 pass the security tests. Admission follows the publisher's functional label, and the table reports our locally verified outcomes.

\begin{table}[t]
\caption{Cohort inventory, locally verified original outcomes, and reference-symbol review coverage. Verif. counts completed original evaluations. Func and Sec refer to that subset. Coverage uses all 387 admitted patches.}
\label{tab:inventory}
\centering\footnotesize
\setlength{\tabcolsep}{3pt}
\renewcommand{\arraystretch}{1.12}
\begin{tabular}{@{}lrrrrr@{}}
\toprule
Coding generator & Source & Admitted & Verified & Func & Sec \\
\midrule
Claude 4 Sonnet & 186 & 88 & 83 & 81 & 15 \\
Gemini 2.5 Pro & 186 & 29 & 24 & 24 & 6 \\
Gemini 3 Pro & 186 & 58 & 55 & 54 & 16 \\
GLM-4.7-Flash & 186 & 38 & 30 & 30 & 5 \\
GLM-5 & 186 & 100 & 98 & 98 & 17 \\
Kimi K2 & 186 & 33 & 28 & 27 & 4 \\
Qwen3-Coder-Next & 186 & 41 & 36 & 36 & 6 \\
\midrule
Total & 1302 & 387 & 354 & 350 & 69 \\
\midrule
\multicolumn{6}{@{}l}{\textit{Reference-symbol review coverage among 387 admitted patches}} \\
\multicolumn{4}{@{}l}{SAST reference file} & 228 & 58.9\% \\
\multicolumn{4}{@{}l}{SAST primary reference symbol} & 52 & \textbf{13.4\%} \\
\multicolumn{4}{@{}l}{SAST symbol including flow endpoints} & 63 & 16.3\% \\
\multicolumn{4}{@{}l}{Change-informed plan reference symbol} & 366 & \textbf{94.6\%} \\
\bottomrule
\end{tabular}

\end{table}

\subsection{Evaluated Configurations and Investigator Models}
The two security investigators are DeepSeek-V3.2, accessed through OpenRouter's SiliconFlow FP8 endpoint, and Gemini-3.8-Flash, accessed through the native Gemini API. Both run at temperature zero with a 22-turn cap and a 100,000-byte message budget, and Gemini uses LOW thinking. Every configuration receives the functional request, the modified symbols, the scanner findings, and the same repair protocol.
\begin{itemize}
\item \textbf{C0 (Snapshot + SAST).} Current source of the modified symbols and static warnings, without historical AST edits or recorded reasoning.
\item \textbf{C1 (C0 + Temporal AST).} Adds the structural edits showing how those symbols evolved.
\item \textbf{C2 (C0 + Recorded Rationale).} Adds the coding agent's original development explanations.
\item \textbf{C3 (C0 + Both).} Combines temporal AST edits and recorded explanations.
\end{itemize}
The seven models in Table~\ref{tab:inventory} are the coding generators that wrote the original patches, and the two investigators review every generator's outputs. C1, C2, and C3 also receive audit cues that C0 does not. Comparisons against C0 therefore evaluate complete evidence bundles, while comparisons among C1, C2, and C3 isolate the contribution of each history channel.

All configurations use Semgrep CE 1.101.0 with the Python rules at commit \code{495df89c}, including correctness and maintainability rules, and CodeQL 2.20.0 with Python queries 1.3.4 and the security-extended suite. Investigation inputs scope warnings to changed files and to at most twelve symbol groups, and coverage audits use the full scanner output. SAST-seeded patches receive at least one scoped warning, and unseeded patches receive none. The benchmark hides the repository's prior Git history, so the history available to the investigators is the coding agent's recorded trajectory after the reference commit.

\subsection{Evaluation Protocol}
We run SusVibes's official functional and security tests inside isolated containers. FuncPass and SecPass require recognized terminal test summaries, including the benchmark's expected-failure allowances. Hidden security tests and reference patches serve only for retrospective evaluation and are never given to an investigator as feedback. Investigators receive public functional feedback.

Each admitted patch is reviewed in four configurations by two investigators, for 3,096 repair runs. Table~\ref{tab:results} compares the 345 patches whose original and candidate evaluations completed in all eight configurations, and DeepSeek additionally supports a comparison over 354 patches. An investigator error can leave a retained patch or a no-op that is evaluated later, in the same restored environment, with results reused only for byte-identical patches. We report conditional repair outcomes rather than benchmark-wide generator scores. Task-cluster intervals use 5,000 bootstrap replicates with seed 12, to account for repeated generator attempts on the same task~\cite{bootstrap}. A security conversion is an originally failing security outcome that becomes a verified pass after repair. Preservation of an original pass is recorded separately. An incomplete evaluation stays unresolved and is not counted as a failed security assertion.

\section{Benchmark Repair Results (RQ1--RQ3)}
\label{sec:results}
This section answers the three benchmark questions on coverage, end-to-end verified repair, and the contribution of each provenance channel.

\subsection{RQ1. Symbol Localization Coverage}
Across the 387 admitted patches, CodeQL and Semgrep warnings fall in a reference security file in 228 cases (58.9\%). The primary warning location falls on a reference symbol in 52 cases (13.4\%), and 63 cases (16.3\%) when flow endpoints are included. These figures measure how often the scanner directs attention to the relevant symbol and should not be read as vulnerability detection recall.

Tracking modified symbols and their relations places a reference symbol in the review scope for 366 patches (94.6\%), with the complete function body in 329. Among the 335 patches with no primary warning on the reference symbol, change tracking still supplies the full body in 282, and the task request names the reference target in 307 of them. A reference-symbol match determines which code to examine. It does not identify the missing condition.

\rqfinding{RQ1}{How often does symbol-level change tracking bring the vulnerable symbol into the review scope, compared with stock SAST locations?}{\tracer{} places the reference symbol in the review scope for 94.6\% of patches, against 13.4\% for the primary SAST location.}

\subsection{RQ2. End-to-End Verified Security Repair}
\begin{table*}[t]
\caption{Verified repair outcomes. Panel A reports FuncPass and SecPass counts. Panel B reports SecPass for DeepSeek and Gemini on the same 345 patches. Bold marks each investigator's maximum across C0 to C3, including ties.}
\label{tab:results}
\centering\footnotesize
\renewcommand{\arraystretch}{1.12}
\begin{tabular*}{\textwidth}{@{\extracolsep{\fill}}lrrrrrr@{}}
\toprule
Population or investigator & $N$ & Original & C0 & C1 & C2 & C3 \\
\midrule
\multicolumn{7}{@{}l}{\textbf{A. End-to-end outcomes. FuncPass / SecPass counts}} \\
DeepSeek-V3.2 (common cohort) & 345 & 342 / 68 & 341 / 71 & 342 / 73 & \textbf{342 / 76} & 342 / 74 \\
Gemini-3.8-Flash (common cohort) & 345 & 342 / 68 & 341 / 70 & 340 / 74 & \textbf{341 / 75} & 340 / 73 \\
DeepSeek-V3.2 (full completed cohort) & 354 & 350 / 69 & 349 / 72 & 350 / 74 & \textbf{350 / 77} & 350 / 75 \\
\midrule
\multicolumn{7}{@{}l}{\textbf{B. Common-cohort SecPass by coding generator and stratum. DeepSeek / Gemini}} \\
Claude 4 Sonnet & 83 & 15 & 16 / 16 & 16 / 16 & \textbf{18} / \textbf{18} & 16 / 16 \\
Gemini 2.5 Pro & 24 & 6 & \textbf{6} / \textbf{6} & \textbf{6} / \textbf{6} & \textbf{6} / \textbf{6} & \textbf{6} / \textbf{6} \\
Gemini 3 Pro & 55 & 16 & 16 / 16 & \textbf{17} / 16 & \textbf{17} / 17 & \textbf{17} / \textbf{18} \\
GLM-4.7-Flash & 30 & 5 & 5 / \textbf{6} & \textbf{6} / \textbf{6} & 5 / 5 & 5 / \textbf{6} \\
GLM-5 & 93 & 16 & 16 / 16 & 16 / 16 & \textbf{18} / \textbf{17} & 17 / 16 \\
Kimi K2 & 26 & 4 & \textbf{6} / 4 & \textbf{6} / \textbf{8} & \textbf{6} / 6 & \textbf{6} / 5 \\
Qwen3-Coder-Next & 34 & 6 & 6 / \textbf{6} & 6 / \textbf{6} & 6 / \textbf{6} & \textbf{7} / \textbf{6} \\
SAST-seeded (Scoped warning) & 176 & 36 & 37 / 38 & 37 / \textbf{40} & \textbf{40} / \textbf{40} & 38 / \textbf{40} \\
Unseeded (No scoped warning) & 169 & 32 & 34 / 32 & \textbf{36} / 34 & \textbf{36} / \textbf{35} & \textbf{36} / 33 \\
Development exposed & 235 & 44 & 45 / 45 & 48 / 47 & \textbf{50} / \textbf{48} & \textbf{50} / \textbf{48} \\
Previously protected & 110 & 24 & \textbf{26} / 25 & 25 / \textbf{27} & \textbf{26} / \textbf{27} & 24 / 25 \\
\bottomrule
\end{tabular*}

\end{table*}
Of the 345 shared patches, 342 initially pass the functional tests and 68 pass the security tests. With DeepSeek as investigator, C0 raises the security count to 71, C1 to 73, and C2 to 76 (22.03\%). Gemini shows the same ordering, with 70 under C0, 74 under C1, and 75 under C2 (21.74\%). C2 adds five net security passes over C0 with each investigator. The comparison is paired on the same original patches, so the difference is attributable to the investigation and repair rather than to a different task mix.

Functional regressions are rare. DeepSeek retains all 342 functional passes under C1, C2, and C3, and loses one under C0. Gemini retains 341 under C0 and C2, and 340 under C1 and C3. The larger DeepSeek comparison over 354 patches keeps the same security ranking, with C2 reaching 77 passes from 69 originals. The two outcomes are reported side by side because an unchanged candidate preserves functionality without repairing the flaw.

\rqfinding{RQ2}{Does guarded investigation improve verified security outcomes without degrading functionality?}{C2 raises verified security passes from 68 to 76 with DeepSeek and to 75 with Gemini, with zero functional regressions for DeepSeek and one for Gemini.}

\subsection{RQ3. Impact of Provenance Channels}
Recorded rationale yields the highest aggregate SecPass count with both investigators. DeepSeek under C2 makes eight conversions across six distinct tasks, and Gemini under C2 makes seven across four tasks. These counts refer to repaired generator outputs, and several conversions concern the same task implemented by different generators.

Combining both channels yields fewer verified repairs than rationale alone. DeepSeek falls from 76 under C2 to 74 under C3, and Gemini from 75 to 73. Context dilution is one possible explanation, but the comparison does not establish the mechanism. Rankings at the generator level are less uniform (Table~\ref{tab:results}), and for C2 against C0 each investigator has six wins and one loss across the seven generators. The task-cluster 95\% intervals for the C2 gain over C0 are 0.00 to 2.96 percentage points for DeepSeek and $-0.32$ to 3.86 for Gemini. The effect is small and both intervals include zero. The data support a consistent ordering across two investigators and do not support a claim of large improvement. Table~\ref{tab:results} also splits the cohort into 235 development-exposed patches and 110 previously protected patches. On the previously protected subset, C2 adds no security passes over C0 for DeepSeek and two for Gemini.

\rqfinding{RQ3}{How do temporal AST history, recorded rationale, and their combination affect verified repair?}{Recorded rationale ranks highest with both investigators. Adding temporal AST history on top of it lowers the verified security count for both.}

\section{Repair Funnel and Failure Analysis (RQ4)}
\label{sec:engineering}
This section follows every repair attempt through four nested steps, from the first hypothesis on a reference symbol to a retained candidate, and examines the three points at which most candidates are lost.

\subsection{The Four-Step Repair Funnel}
High review coverage does not guarantee a correct diagnosis or a correct repair. Table~\ref{tab:engineering} reports, for each investigator and configuration, how many of the 387 admitted patches reach each of four nested steps.

\begin{table*}[t]
\caption{Four-step investigation and repair funnel across 387 admitted patches per configuration. SecPass and new fixes count only candidates within Step 4. Bold highlights supported assessments and C2's new fixes.}
\label{tab:engineering}
\centering\footnotesize
\setlength{\tabcolsep}{4pt}
\renewcommand{\arraystretch}{1.12}
\begin{tabular*}{\textwidth}{@{\extracolsep{\fill}}lcccccc@{}}
\toprule
Investigator & Config. & \shortstack{Step 1: Hypothesis\\on Target Symbol} & \shortstack{\textbf{Step 2: Supported}\\\textbf{Assessment}} & \shortstack{Step 3: Reference-File\\Edit} & \shortstack{Step 4: Nonempty\\Retained Patch} & \shortstack{Verified SecPass\\(New fixes)} \\
\midrule
DeepSeek-V3.2 & C0 & 245 & \textbf{109} & 89 & 58 & 13 (4 new) \\
DeepSeek-V3.2 & C1 & 251 & \textbf{109} & 98 & 65 & 13 (5 new) \\
DeepSeek-V3.2 & C2 & 237 & \textbf{85} & 74 & 54 & \textbf{13 (8 new)} \\
DeepSeek-V3.2 & C3 & 252 & \textbf{97} & 89 & 60 & 10 (6 new) \\
\midrule
Gemini-3.8-Flash & C0 & 213 & \textbf{15} & 14 & 11 & 3 (2 new) \\
Gemini-3.8-Flash & C1 & 225 & \textbf{32} & 28 & 20 & 8 (6 new) \\
Gemini-3.8-Flash & C2 & 223 & \textbf{29} & 27 & 18 & \textbf{8 (7 new)} \\
Gemini-3.8-Flash & C3 & 235 & \textbf{23} & 21 & 17 & 5 (4 new) \\
\bottomrule
\end{tabular*}

\end{table*}

\begin{enumerate}
\item \textbf{Hypothesis on Target Symbol.} The investigator records a hypothesis on a symbol touched by the reference security patch (237 to 252 cases per configuration for DeepSeek, 213 to 235 for Gemini).
\item \textbf{Supported Assessment.} The investigator records an accepted supported assessment on a reference symbol (85 to 109 for DeepSeek, 15 to 32 for Gemini). This step records the investigator's own assessment and does not independently confirm the flaw.
\item \textbf{Reference-File Edit.} The case contains an edit in a reference file, which need not address the same symbol or hypothesis (74 to 98 for DeepSeek, 14 to 28 for Gemini).
\item \textbf{Nonempty Retained Patch.} The case ends with a nonempty candidate patch (54 to 65 for DeepSeek, 11 to 20 for Gemini). Retention is a protocol outcome and carries no functional or security verdict. The final column reports official security outcomes for this subset.
\end{enumerate}

Each step includes only the cases that reached the preceding steps, so the table shows at which step a candidate leaves the workflow. It answers a different question from Table~\ref{tab:results}, which reports the final outcome of every evaluated patch, including those left unchanged.

\subsection{Drop-off 1. False Refutations During Diagnosis}
The largest loss occurs between forming a hypothesis and recording support for it. The ledgers show premature refutations, hypotheses on the wrong property, inconclusive assessments, and turn exhaustion. In several cases the investigator inspects a partial check or a helper call and concludes that the boundary is already protected. In one attribute-access investigation, for example, the existing restriction on underscore-prefixed names is judged sufficient, although inspection attributes remain unrestricted. In other cases the investigator finds the correct function and then examines an unrelated property, for instance dictionary copying while the actual contract concerns expiry. Because the funnel also includes outputs that were secure to begin with, an accepted refutation is not necessarily a false all-clear. These examples show that locating the relevant code and understanding its security obligation are distinct steps. A guard can be present and still enforce a property other than the one required at the boundary.

\subsection{Drop-off 2. Turn Exhaustion Before Validation}
A late edit can exhaust the turn budget before public validation completes, and validation slots and required probe reruns can also prevent retention. A separate recovery check, run during development, evaluates seventy saved candidates that were discarded for protocol completion reasons. Of these, 68 reached a turn or validation-slot limit and two lacked a probe rerun. Offline tests find 39 functional passes, eighteen failures, and thirteen unresolved executions. Nine candidates pass the security tests, eight of them preserved successes and one a new conversion. These recovery results are kept separate from the campaign scores. They indicate that a validation opportunity reserved after the final edit would retain usable candidates, without establishing that every loss was due to scheduling.

\subsection{Drop-off 3. Incomplete Fixes and Contract Mismatches}
\begin{table}[t]
\caption{Manual inspection of fifteen retained patches with reference-symbol edits, verified functional passes, and failed security tests. Categories describe this selected diagnostic sample.}
\label{tab:taxonomy}
\centering\footnotesize
\begin{tabularx}{\columnwidth}{@{}>{\raggedright\arraybackslash}p{0.25\columnwidth}r>{\raggedright\arraybackslash}X@{}}
\toprule
Failure category & Count & Repair attempt and remaining defect \\
\midrule
\textbf{Right bug class, incomplete fix} & \textbf{9} & \textbf{A plausible fix leaves a bypass or incomplete boundary.} Path normalization misses symlink escape. Size checks follow decompression. Connection validation skips inherited settings. Request checks miss framing variants. URL checks leave a backslash bypass. \\
\addlinespace
Wrong security property & 4 & Session checks concern keys rather than expiry. Request checks limit body size rather than media-type handling. Other candidates leave route anchoring or username normalization unresolved. \\
\addlinespace
Right condition, wrong failure behavior & 2 & An empty axis raises \code{IndexError} instead of \code{ValueError}. Malformed-byte decoding still raises instead of following the required failure behavior. \\
\bottomrule
\end{tabularx}

\end{table}
A retained patch that passes the functional tests can still leave the security contract unsatisfied. We inspected fifteen such cases, outside the diagnostic pilot, by comparing the candidate diff, the reference patch, and the official failing assertions. Fourteen come from DeepSeek and one from Gemini. Table~\ref{tab:taxonomy} classifies nine of the fifteen as incomplete repairs of the correct vulnerability class, such as lexical path normalization that does not account for symbolic links, a size check applied after decompression, and validation that omits inherited connection settings. Four patches address a security property other than the one tested, such as session-key handling instead of expiry, or body-size limits instead of media-type parsing. Two patches enforce the right condition with the wrong failure behavior. In all three categories, verifying the repair requires checking the full input and error contract rather than the presence of a new guard. A size check applied after decompression cannot prevent resource consumption that has already occurred, and the repair must enforce the intended boundary at the point where the unsafe operation can still be prevented.

\subsection{Operational and Provider Failures}
Gemini records 205 repair errors, comprising 153 \code{blockReason: OTHER} responses, 45 JSON decoding errors, six HTTP 500 errors, and one other value error. DeepSeek records 107 errors, comprising 63 explicit stop interruptions, 32 HTTP 429 responses, nine deadlines, two value errors, and one connection error. The recorded labels distinguish provider responses, parsing failures, and run interruptions, and we do not attribute causes beyond what the labels state. Final candidate evaluations are reported separately from these investigation errors.

\rqfinding{RQ4}{Where are candidates lost between symbol localization and verified repair?}{Candidates are lost during diagnosis, during validation scheduling, and in semantic repair. The funnel separates these losses from operational errors and from the final test outcome.}

\section{Industrial Case Study. Structural Biology Pipelines (RQ5)}
\label{sec:practice}
This section evaluates the maintenance stage of \tracer{} on software outside the SusVibes benchmark. Following the silent-fix setting of CoLeFunDa~\cite{colefunda}, we repair five components of structural-biology software selected with our industrial co-author, then submit ten feature requests against the repaired code and measure, in a $2\times2$ factorial, whether repair memory and boundary feedback allow the feature to be delivered without undoing the repair.

\begin{table*}[t]
\caption{Independent biology maintenance: two tasks per repository. Rows report final FuncPass/SecPass/JointPass after at most one retry. FuncPass combines ordinary and feature checks. SecPass combines saved and extended checks, including DockQ and AutoDock-Vina reliability contracts. Aggregate counts are out of ten. Bold highlights B3 (Both).}
\label{tab:bio}
\centering\footnotesize
\setlength{\tabcolsep}{4pt}
\renewcommand{\arraystretch}{1.12}
\begin{tabularx}{\textwidth}{@{}l >{\raggedright\arraybackslash}X c c c c@{}}
\toprule
Repository & Two feature tasks & B0 & B1 & B2 & \textbf{B3}\\
 & & Functional & Memory & Boundary & \textbf{Both}\\
\midrule
DockQ & Lazy model iteration, parsed-model cache & 1/1/1 & 2/2/2 & 1/2/1 & \textbf{2/2/2}\\
haddock3 & Internal TAR aliases, selected nested extraction & 1/2/1 & 1/2/1 & 0/2/0 & \textbf{2/2/2}\\
ProteinMPNN & Hierarchical names and metadata, name mapping & 2/0/0 & 2/1/1 & 2/0/0 & \textbf{2/1/1}\\
RFdiffusion & Legacy config retry, checkpoint bytes cache & 1/1/0 & 2/2/2 & 1/1/0 & \textbf{2/2/2}\\
AutoDock-Vina & Dry-run preview, pose text in memory & 2/2/2 & 2/2/2 & 2/2/2 & \textbf{2/2/2}\\
\midrule
\multicolumn{2}{@{}l}{\textbf{First proposal, total (out of 10)}} & 5/6/4 & 6/6/5 & 4/7/3 & \textbf{6/6/5}\\
\multicolumn{2}{@{}l}{\textbf{Final proposal, total (out of 10)}} & 7/6/4 & 9/9/8 & 6/7/3 & \textbf{10/9/9}\\
\multicolumn{2}{@{}l}{Final boundary assertion violations / unknowns} & 3/1 & 1/0 & 3/0 & \textbf{1/0}\\
\multicolumn{2}{@{}l}{Accepted proposals with observed boundary violation} & 3 & 1 & 0 & \textbf{0}\\
\multicolumn{2}{@{}l}{Fallbacks (not feature successes)} & 3 & 1 & 7 & \textbf{1}\\
\bottomrule
\end{tabularx}

\end{table*}

\subsection{Repair Contracts and Independent Scope}
With our co-author at Nostrum Biodiscovery, we selected six open-source tools relevant to the company's pipeline. RFdiffusion~\cite{rfdiffusion}, ProteinMPNN~\cite{proteinmpnn}, haddock3~\cite{haddock3}, DockQ~\cite{dockq}, AutoDock-Vina~\cite{vina}, and foldseek~\cite{foldseek} cover protein structure generation, sequence design, docking, docking quality assessment, and structure search. Semgrep 1.101.0 reports 420 warnings across the six repositories. Review of source and history identifies five Python repair targets, none of which coincides with a warning location. Foldseek contributes eleven warnings but no suitable project-owned Python case and is therefore not part of the maintenance study. We verified the following five component contracts.
\begin{enumerate}[label=\arabic*.]
\item \textbf{DockQ (\code{load_PDB}).} A parser failure can leave a file handle open before a fallback parser opens another. The repair closes ordinary and compressed handles on both success and failure.
\item \textbf{haddock3 (\code{extract_files_flat}).} A flattened archive name can follow a pre-existing destination symlink to a location outside the extraction directory. The repair resolves paths, checks containment, and excludes non-regular archive members.
\item \textbf{ProteinMPNN.} \code{StructureDataset.__init__} accepts unvalidated JSONL record names that later form part of output paths. The repair accepts only nonempty simple filenames and preserves ordinary identifiers and filtering.
\item \textbf{RFdiffusion (\code{Sampler.load_checkpoint}).} Unrestricted pickle loading under PyTorch 2.5.1 permits code execution from an untrusted checkpoint~\cite{pytorchserialization,pythonpickle}. The repair sets \code{weights_only=True} with a scoped \code{safe_globals([slice])} allowlist for the accepted checkpoint forms.
\item \textbf{AutoDock-Vina (\code{Vina.write_poses}).} Numeric checks are skipped on entry to an \code{elif not overwrite} branch. The repair validates pose counts and energy ranges irrespective of the overwrite setting.
\end{enumerate}
The DockQ and AutoDock-Vina cases are reliability contracts and not exploitable vulnerabilities. Four of the five repairs were written by the model. For RFdiffusion, two model attempts failed and the remedy was specified by the developer. All five boundary assertions fail before repair and pass after it, and ordinary-input controls pass in both states. An external note for each repair records the invariant, its rationale, the source identity, and the check. Before the maintenance tasks, we strip comments and docstrings with \code{tokenize} and \code{ast}, keep the executable AST unchanged, and re-verify the cleaned starting code~\cite{pythontokenize,pythonast}, so that the model receives no textual hint of the repair.

\subsection{Equal-Budget Feature Evolution}
The ten tasks, two per repository, extend checkpoint compatibility, archive aliases, hierarchical names, in-memory pose output, and lazy parsing (Table~\ref{tab:bio}). They form a purposive stress sample, independent of SusVibes, across five repository clusters. The checks cover both the default mode and the newly requested mode, and each repaired starting point passes its existing checks. For every task, an evaluator-written reference implementation passes all checks and a deliberately weakened reference delivers the feature but violates a boundary. Both references are withheld from the model and serve only to validate the checks.

DeepSeek-V3.2 at temperature zero receives the cleaned source and the feature request. The factorial varies repair memory and boundary feedback. Memory supplies the repair note together with a prospective clarification of the contract. Boundary feedback adds the saved and extended checks to the feedback the model receives. All four conditions receive functional feedback, the same token budget, and at most one retry. In B0 and B1, the boundary checks are run after generation and do not influence the retry. All 40 cells are retained, checks run in pinned offline containers, and the archive preserves the protocol and the correction history. FuncPass requires the ordinary and feature checks to pass, SecPass requires the saved and extended boundary checks to pass, and JointPass requires both. Table~\ref{tab:bio} reports first and final proposals separately. An execution error is recorded as an unknown boundary outcome rather than as a violation. A fallback retains the repaired starting code and is counted as a failure to deliver the feature.

\noindent\textbf{Functional feedback alone (B0).} B0 reaches 4/10 joint successes and accepts three proposals that violate a contract. The RFdiffusion proposal re-enables unrestricted checkpoint loading with \code{weights_only=False}, as described in Section~\ref{sec:motivation}. Two ProteinMPNN proposals admit backslashes that the portable naming contract forbids, although neither demonstrates path traversal on Linux. The functional checks accept all three because the requested feature works.

\noindent\textbf{Boundary gate alone (B2).} B2 accepts no observed violation but reaches only 3/10 joint successes, with seven fallbacks. Without the repair note, the model receives the failing check but has no record of why the boundary exists. The gate protects the boundary, and that protection is not sufficient to produce a proposal that satisfies both obligations.

\noindent\textbf{Memory alone (B1).} With the note, the model implements the feature within the repaired contract in 8/10 tasks. One accepted proposal still relaxes ProteinMPNN's default simple-name restriction. Functional checks cannot reject this naming violation, even though the note states the intended contract.

\noindent\textbf{Memory and boundary feedback (B3).} With both, 9/10 tasks succeed jointly and no contract violation is accepted. The gate rejects the remaining naming violation and retains the repaired starting code. The note tells the model how to preserve the boundary, and the executable check determines whether the proposal does so. The five joint successes that B3 adds over B0 consist of one restricted-loading case, one portable-naming case, and three feature or execution fixes. B3 leads in this sample, and five selected repository clusters do not establish a population effect.

\rqfinding{RQ5}{Can repair memory and boundary gates preserve repaired properties while subsequent feature changes succeed?}{Under the same one-retry limit, joint success is 4/10 with functional feedback alone, 8/10 with memory, 3/10 with a boundary gate, and 9/10 with both. B3 rejects its one remaining contract-violating proposal. A fallback is counted as a failure to deliver the feature.}

\section{Industrial Relevance and Practical Takeaways}
\label{sec:implications}
This section discusses the implications of the two studies for teams that integrate coding agents into their development process and summarizes the lessons we consider transferable to other pipelines.

\subsection{Relevance to Industrial Software Engineering}
In most industrial settings a change is accepted when the functional test suite passes, and our results indicate that this criterion is not sufficient for security. In the SusVibes cohort, 350 of the 354 patches with completed evaluations pass their functional tests, whereas 69 pass the associated security tests. In the structural-biology software, the five repaired components all read data from outside the program, and in each case the defect was located in code that passed the project's existing tests. In both settings, the missing condition could be traced to an assumption about an upstream caller, a file path, or a serialization format that was never made explicit in the submitted code. Static analysis remains the first line of defense, and we do not propose to replace it. We observe, however, that in our cohort the primary warning location of CodeQL and Semgrep coincides with the symbol modified by the reference security fix in 13.4\% of the admitted patches. \tracer{} is intended to complement static analysis at two points of a CI/CD workflow. During review, it uses the development history and the recorded reasoning of the coding agent to direct attention to the symbols that the pull request modified and to the assumptions that the agent stated about them. After a security fix has been accepted, it records the fix as a JSON repair note together with an executable boundary check, so that subsequent feature updates to the same component are tested against the earlier security work.

\subsection{Actionable Takeaways for Practitioner Pipelines}
\noindent\textbf{1) More historical context is not always better.}
In our benchmark study, the configuration that adds the agent's recorded rationale to the current source and scanner findings (C2) produced more verified security passes than the configuration that adds both the rationale and the temporal AST edits (C3), with 76 against 74 passes for DeepSeek and 75 against 73 for Gemini. Our experiment does not identify the cause of this difference. Context dilution is a plausible explanation, since the investigator receives a larger and less focused prompt in C3, but it is not established by our data. The rationale provides a specific claim about the code, for example that an upstream caller already sanitizes its input, which the investigator can confirm or refute against the current implementation. For teams that log agent trajectories, this suggests storing reasoning excerpts linked to the symbols they modified and retrieving at review time only the excerpts that concern the symbol under investigation.

\noindent\textbf{2) Combine change tracking with static analysis.}
Symbol-level change tracking placed the symbol modified by the reference security fix in the review scope for 94.6\% of the admitted patches, compared with 13.4\% for the primary scanner location, and among the 335 patches for which the scanner did not point at that symbol it supplied the complete body in 282 cases. We therefore recommend that a review workflow for agent-generated changes define its scope from the modified symbols and their callers, and then use scanner warnings and recorded assumptions as evidence within that scope, so that a modified function without a warning can also be investigated. Coverage of the relevant symbol is a precondition for diagnosis and does not by itself indicate whether a condition is missing.

\noindent\textbf{3) Reserve validation capacity after the final edit.}
During protocol development we evaluated seventy candidate patches that the campaign had discarded, 68 of them because the turn budget or the validation slot had been exhausted (Section~\ref{sec:engineering}). Of these candidates, 39 passed the functional tests and nine passed the security tests, including one new conversion. These candidates were lost to scheduling and not to the quality of the repair. We conclude that a validation step should be reserved after the last edit, independently of the budget consumed by exploratory probing. Recording candidate generation, retention, and completed evaluation as separate events also allowed us to recover from the 205 and 107 provider or protocol errors of the two investigators without repeating model requests.

\noindent\textbf{4) Separate repair guidance from acceptance checks.}
The factorial design of the maintenance study allows the two interventions to be assessed separately. Repair memory alone led to joint success in 8 of 10 tasks but accepted one proposal that relaxed the naming contract repaired earlier. A boundary gate alone accepted no violating proposal but led to joint success in only 3 of 10 tasks. The combination led to joint success in 9 of 10 tasks with no accepted violation, whereas functional feedback alone led to 4 of 10 and accepted three violating proposals. The note informs the model of the invariant and the reason for it, which allows the feature to be implemented within the contract, and the check determines whether a given proposal respects it. We therefore recommend that teams store both with each accepted security fix. The check should cover the modes introduced by the new feature as well as the default mode, since in the ProteinMPNN tasks the proposals that violated the naming contract did so only on the new naming path. When a feature request widens the input domain of a repaired component, the contract should be revised explicitly, and a fallback that retains the repaired code should be reported as a feature that was not delivered rather than as a successful outcome.

\section{Threats to Validity}
\label{sec:threats}
\emph{Construct validity.} Overlap with the reference patch measures the opportunity to review the relevant code, and a broad symbol or an incidental edit can overstate the precision of a diagnosis. The SusVibes tests verify the specified contract and do not certify the absence of other vulnerabilities~\cite{susvibes,susvibesartifact,repairsurvey,overfitting}. The recorded rationale is a stated development trace and is not guaranteed to reflect the model's internal reasoning~\cite{turpin,lanham}.

\emph{Internal validity.} The protocol was developed through repeated inspection of failures, so that individual engineering decisions were not evaluated as isolated interventions. The fifteen semantic reviews form a purposive sample inspected by a single reviewer and do not support a population estimate. The maintenance factorial holds the retry ceiling fixed but uses a single investigator and a single continuation per cell, and the memory condition includes prospective clarifications of the contract. The feature tasks were designed with knowledge of the repairs and constitute a stress sample.

\emph{External validity.} Selection favors Python outputs with publisher-reported functional success. The 42 admitted patches without complete shared outcomes may differ from the rest in environment difficulty or provider reliability. Repeated tasks across generators and the small number of security conversions limit precision. The scientific-software study covers ten tasks clustered within five selected components and was run by the researchers with input from the industrial co-author. Portable naming restrictions and reliability contracts do not by themselves constitute exploitable vulnerabilities. Our results do not establish the effectiveness of the approach for other languages or in production deployments.

\section{Related Work}
\label{sec:related}
Our work relates to four lines of research, reviewed in turn below.

\subsection{Benchmarks and Security of Agent-Generated Code}
SWE-bench evaluates the resolution of repository issues~\cite{swebench}, and SusVibes separates functional from security outcomes for agent-generated changes~\cite{susvibes,susvibesartifact}. CVEfixes links historical vulnerabilities to their fixing commits, and Vul4J provides reproducible vulnerabilities with triggering tests~\cite{cvefixes,vul4j}. SWE-agent defines the interface through which an agent interacts with a repository~\cite{sweagent}, and ReAct interleaves explanations with actions~\cite{react}. We use the record these interfaces preserve as investigation evidence and evaluate existing generated patches against the security oracle of the benchmark, rather than generating new patches.

\subsection{Static Analysis and LLM-Assisted Vulnerability Detection}
Beyond CodeQL and Semgrep~\cite{codeql,semgrep}, IRIS infers taint specifications with an LLM~\cite{iris}, LLMDFA combines model reasoning with dataflow checks~\cite{llmdfa}, RepoAudit explores repository dataflows with memory and validation~\cite{repoaudit}, PAGENT guides the generation of vulnerability-triggering inputs with program analysis~\cite{pagent}, and KNighter synthesizes checkers from historical patches~\cite{knighter}. None of these approaches uses the development trajectory and recorded rationale of the coding agent that produced the code as evidence. \tracer{} compares these evidence channels within a single guarded repair protocol, and its conditional runtime checks follow PAGENT in grounding a suspected flaw in observed program behavior before a repair is attempted.

\subsection{Code Evolution, Silent Fixes, and Automated Program Repair}
GumTree computes structural edit scripts between versions of a source file~\cite{gumtree,gumtree3}, ViolationTracker tracks static-analysis warnings across revisions~\cite{violationtracker}, CoLeFunDa represents function-level changes in order to identify silent vulnerability fixes~\cite{colefunda}, and FixMiner mines recurring repair patterns~\cite{fixminer}. On the repair side, InferFix uses analyzer guidance~\cite{inferfix}, ChatRepair uses conversational test feedback~\cite{chatrepair}, and RepairAgent inspects, edits, and tests code autonomously~\cite{repairagent}. \tracer{} adopts the function-level change representation of CoLeFunDa and adds an analysis of where candidate repairs are lost between diagnosis, retention, and verified repair.

\subsection{Persistent Memory and Continuous Software Evolution}
Reflexion stores verbal feedback in an episodic memory~\cite{reflexion}, and Self-Refine iteratively revises its own outputs~\cite{selfrefine}. POSTURA maintains a graph of the security posture of a system~\cite{postura}, and \emph{Undoing the Fix} studies deliberately elicited security regressions in an AI-control setting~\cite{undoingfix}. \tracer{} pairs each repair note with an independently owned executable check and evaluates benign feature evolution, rather than adversarial regression, against the saved contract.

\section{Conclusion}
We have presented \tracer{}, which uses the development history of an agent-generated change as evidence for security investigation and preserves the result of that investigation as a contract for subsequent maintenance. On SusVibes, symbol-level tracking placed the reference vulnerable symbol in the review scope for 94.6\% of patches, compared with 13.4\% for the primary SAST location, and the configuration that includes the recorded rationale produced the most verified security passes with both investigators, with at most one functional regression. In the structural-biology maintenance study, repair memory combined with boundary feedback delivered the requested feature while preserving the repair in 9 of 10 tasks, against 4 of 10 for functional feedback alone, and rejected the only remaining proposal that violated a contract.

\section*{Data Availability}
The skill, analysis helpers, derived outcome records, supporting tables, and the RQ5 prompts, responses, candidates, and checks are archived on Zenodo~\cite{tracerartifact}. Benchmark inputs originate from SusVibes~\cite{susvibesartifact}. Raw transcripts, evaluator logs, and container exports are excluded.

\section*{Acknowledgements}
We thank Google for Gemini API credits and Nostrum Biodiscovery for verification assistance and feedback.

\newcommand{\BIBdecl}{\interlinepenalty=10000}
\balance
\bibliographystyle{IEEEtran}
\bibliography{references}
\end{document}